\documentclass[twocolumn,showpacs,preprintnumbers,amsmath,amssymb,eqsecnum]{revtex4}

\usepackage{graphicx, subfigure}
\usepackage{dcolumn}
\usepackage{bm}
\usepackage{epsfig}
\usepackage{verbatim}

\def\z{{\mathbf{z}}}

\begin{document}


\title{The Leggett-Garg Inequalities and Non-Invasive Measurability}

\author{JM Yearsley}


\affiliation{Department of Psychology, School of Arts and Social Sciences, City University London, London EC1R 0JD, UK}

\date{\today}



\begin{abstract}
The Leggett-Garg inequalities are a set of inequalities obeyed by classical systems but violated in quantum theory. Their violation has been taken as evidence that quantum theory lacks a `realistic' formulation. However in addition to realism the derivation of the Leggett-Garg inequalities relies on another, more obscure assumption of `non-invasive measurability.' The significance of this assumption and the consequences for the interpretation of violations of the Leggett-Garg inequalities have been hotly debated. In this paper we present a pedagogical introduction to the issues, focussing on the significance of non-invasive measurability. We give a simple derivation of the Leggett-Garg inequalities paying particular attention to where and when the two assumptions are used, and we give an example of a realist but not non-invasively measurable hidden variable theory which violates them.  We also discuss recent attempts to experimentally implement non-invasive measurements via so called `undetectable measurements'  and we show that even undetectable measurements are not necessarily non-invasive.
\end{abstract}

\maketitle

\newcommand\beq{\begin{equation}}
\newcommand\eeq{\end{equation}}
\newcommand\bea{\begin{eqnarray}}
\newcommand\eea{\end{eqnarray}}
\newcommand\bi{\begin{itemize}}
\newcommand\ei{\end{itemize}}

\def\A{{\cal A}}
\def\D{\Delta}
\def\H{{\cal H}}
\def\E{{\cal E}}
\def\De{{\mathcal{D}}}
\def\p{\partial}
\def\la{\langle}
\def\ra{\rangle}
\def\ria{\rightarrow}
\def\Z{{\bf z}}
\def\t{{\tau}}
\def\d{{\delta}}
\def\y{{\bf y}}
\def\k{{\bf k}}
\def\q{{\bf q}}
\def\p{{\bf p}}
\def\P{{\bf P}}
\def\r{{\bf r}}
\def\de{{\partial}}
\def\s{{\sigma}}
\def\a{\alpha}
\def\b{\beta}
\def\e{\epsilon}
\def\z{\xi}
\def\l{\lambda}
\def\v{\nu}
\def\g{{\gamma}}
\def\G{\Gamma}
\def\w{{\omega}}
\def\Tr{{\rm Tr}}
\def\iff{{\rm iff}}
\def\erf{{\rm Erf}}
\def\ih{{ \frac {i} { \hbar} }}
\def\trho{{\rho}}
\newcommand\bra[1]{\left<#1\right|}
\newcommand\ket[1]{\left|#1\right>}
\newcommand\brak[2]{\left<#1|#2\right>}
\newcommand\but[1]{\ket{#1}\bra{#1}}
\newcommand\ex[1]{\left<#1\right>}
\newcommand\vvec[2]{ \begin{bmatrix} 
      #1  \\
        #2\\
   \end{bmatrix}}

\def\au{{\underline \alpha}}
\def\bu{{\underline \beta}}
\def\pp{{\prime\prime}}
\def\id{{1 \!\! 1 }}
\def\half{\frac {1} {2}}

\section{Introduction}

There has been renewed interest recently in the perennial question of whether quantum theory can be viewed as realist.  That is, can the predictions of quantum mechanics for the outcomes of measurement, which have a necessarily statistical character, be considered as arising from the definite, not necessarily deterministic, behaviour of a real particle etc. about which quantum theory does not give us precise information? This question has a long and interesting history, going back to the origins of quantum theory itself \cite{EPR, Jammer}. 

Recent attention has focussed on the question of whether a single system evolving in time can be given a realist interpretation, in contrast to the usual set up of EPR or the Bell inequalities \cite{EPR,Bell}. 
In a classic paper \cite{LG} Leggett and Garg showed that the predictions of quantum theory for the outcomes of measurement performed on a simple, effectively two level system, were incompatable with two properties which they deemed essential for a realist interpretation. These properties are,
\bi
\item {\em Realism per se:} A system with two or more states will at all times be in one of these states.
\item {\em Noninvasive Measurability:} It is possible, at least in principle, to measure the state of a system without affecting its future evolution.
\ei
Using these two assumptions Leggett and Garg derived an inequality, similar in form the the CHSH inequalities, and showed that it is violated by a series of measurements performed on a simple two level quantum system \footnote{In what follows we shall refer to the inequalities derived in \cite{LG} as `Leggett-Garg inequalities'. They are also commonly referred to as `Temporal Bell inequalities', but we feel this name is misleading for a number of reasons and we shall avoid this terminology.}. 

Ever since the publication of \cite{LG} there has been lively debate in the literature about the significance of the result \cite{LGextra, LGextraextra}. Leaving aside concerns about the particular physical set up envisaged in \cite{LG} the main point of dissagreement has been over the significance of the condition of `noninvasive measurablity.' Some authors have claimed that the predicted violation of the inequality derived in \cite{LG} is entirely due to the failure of this condition to hold in quantum theory. On the other hand Leggett and Garg have argued essentially that (Macroscopic) Noninvasive Measurability is in fact a corollary of the assumption of (Macroscopic) Realism. The resolution of this issue has become more pressing in recent years, as experimental results are now being reported which show direct violation of the Leggett-Garg inequalites \cite{Ex}. In particular in a recent paper \cite{Ex2} violation of a Leggett-Garg inequality was reported using so-called `undetectable measurements.' The claim was that the nature of these measurements meant that the condition of noninvasive measurability was satisfied exactly, so that violation of the Leggett-Garg inequality in this case was a direct refutation of realism.

Our aim in this paper is firstly to provide a pedagogical introduction to the Leggett-Garg inequalities by means of a simple derivation of them, paying particular attention to where and when the assumption of `non-invasive measurability' is used. We will then introduce a simple hidden variable model for a two state system which is explicitly realist but not non-invasively measurable, and we will show that this model can reproduce the results of quantum theory including the predicted violations of the Leggett-Garg inequalities.  This model makes it clear that in general non-invasive measurability is not a corollary of realism and provides evidence to support the view that all experimental tests of the Leggett-Garg inequalities test is the former. Finally we will argue against the view that non-invasive measurability is a property of a particular meaurement set up that can be experimentally determined. This is essentially because the invasiveness or otherwise of a particular measurement is not something that follows from quantum theory, rather it requires reference to a particular theory of hidden variables.

We hope that this paper will be useful to researchers and students in quantum theory unfamiliar with the Leggett-Garg inequalities and also to researchers outside the field of physics interested in applying the tools of quantum theory (see for example \cite{Psych,BuBr}.) 

\section{The Leggett-Garg Inequalities}

In their seminal paper \cite{LG}, Leggett and Garg did not provide a derivation of their inequality. It is clear from their remarks that they believed the derivation was straightforward and that it followed along roughly the same lines as Bell's original derivation of his famous inequality \cite{Bell}. However the lack of an explicit derivation in the original work has tended to increase the confusion in the literature about the exact role of the two assumptions. This has been addressed to some extent by recent work attempting to provide a clearer derivation of the inequalities. The contributions of Maroney \cite{Maroney} have been particularly useful in this area. Nevertheless we feel it is useful, particularly to those unfamiliar with the existing literature, to give as simple and self-contained an account of the derivation as possible. 

Let us begin with the assumption of realism per se. This implies that the expectation value of any Heisenberg picture operator in quantum theory can be written as an average over a set of so-called `hidden variables' $\{\l\}$. For the present discussion it is helpful to think of the hidden variables as essentially the set of possible `trajectories' of the system. Then the observables in quantum theory may be thought of as functions (which also depend on time) from the hidden variables to the set of real numbers $A(\l,t)$. The role of the initial state is to provide a probability distribution on the set of hidden variables which we denote $\rho(\l)$ and which is often referred to as the {\em ontic state}. Then the expectation value of $\hat A(t)$ may be written as,
\beq
\ex{\hat A(t)}=\int d\l A(\l,t) \rho(\l).
\eeq
(In what follows we will explicitly write hats on operators in order to distinguish them from the corresponding functions on the hidden variables.)

So far this is simple because we are only considering a single time. The story becomes more complicated if we consider the expectation value of two or more measurements performed at different times. This is because in general a measurement performed at a time $t$ will disturb the future evolution of the system. The easiest way to keep track of this is to let the probability distribution on the set of hidden variables also depend on the set of measurements performed on the system, so that, $\rho(\l|A,t_{1},B,t_{2}...)$ is the probability distribution on the set of hidden variables $\l$ given the initial state and also that measurement $A$ was carried out at time $t_{1}$, measurement $B$ at time $t_{2}$ etc. Then the correlation function between the outcome of a measurement of $A$ at $t_{1}$ and a measurement of $B$ at $t_{2}$ can be written in this hidden variables model as,
\beq
\ex{\hat B(t_{2})\hat A(t_{1})}=\int d\l B(\l,t_{2}) A(\l,t_{1})\rho(\l|A,t_{1})
\eeq
Note that we can drop the dependence of $\rho$ on the fact that a measurement was performed at $t_{2}$, since there are no further measurement performed after this time so the change in the distribution of $\l$'s caused by this measurement has no consequence. In general we can always ignore the effect of the final measurement on the hidden variables.

Now we can state mathematically the assumption of non-invasive measurability. It is essentially that,
\beq
\rho(\l|A,t_{1},B,t_{2}...)=\rho(\l)\label{2.3}
\eeq
which is obviously interpreted as the assumption that the measurements performed do not change the distribution of $\l$ \footnote{Note that this is essentially the analogue of the locality condition in the treatment of the usual spatial Bell inequalities.}. Note that there is a close relationship between non-invasive measurability and commutativity. Let $t_{-}$ and $t_{+}$ denote two times infinitesimally earlier and later than $t$ respectively. Then the difference between measuring $\hat A$ at $t_{-}$ followed by $\hat B$ at $t_{+}$ compared with the opposite ordering can be written as,
\begin{multline}
\ex{\hat B(t_{+})\hat A(t_{-})-\hat A(t_{+})\hat B(t_{-})}\\=\int d\l B(\l,t_{+}) A(\l,t_{-})\rho(\l|A,t_{-})\\
-\int d\l A(\l,t_{+}) B(\l,t_{-})\rho(\l|B,t_{-})
\end{multline}
With some modest assumptions about continuity this can be simplified to,
\begin{multline}
\ex{\hat B(t)\hat A(t)-\hat A(t)\hat B(t)}\\=\int d\l B(\l,t) A(\l,t)(\rho(\l|A,t)-\rho(\l|B,t))
\end{multline}
$\hat A$ and $\hat B$ therefore commute in this framework if the measurements are non-invasive, or more generally if the measurements cause the same disturbance to the hidden variables. 

We are now in a position to present a simple derivation of one of the Leggett-Garg inequalities. We follow closely the derivation of the Bell inequalities presented in \cite{Bell}. We we consider a dichotomous observable $\hat Q(t)$ with eigenvalues $\pm1$, which is positive if the system is in state 1 and negative if it is in state 2 (the physical significance of the states is not important to this derivation.) $Q(\l,t)$ takes values $\{-1,1\}$ for each $\l$ and $t$. Consider the following quantity,
\begin{multline}
\ex{\hat Q(t_{2})\hat Q(t_{1})}-\ex{\hat Q(t_{4})\hat Q(t_{1})}\\=\int  d\l \left[Q(\l,t_{2})Q(\l,t_{1})-Q(\l,t_{4})Q(\l,t_{1})\right]\rho(\l|Q,t_{1})
\end{multline}
where we have used the assumption of realism to write expectation values in terms of hidden variables. We can rewrite this as,
\begin{multline}
\ex{\hat Q(t_{2})\hat Q(t_{1})}-\ex{\hat Q(t_{4})\hat Q(t_{1})}\\=\int  d\l Q(\l,t_{2})Q(\l,t_{1})\left[1\pm Q(\l,t_{4})Q(\l,t_{3})\right]\rho(\l|Q,t_{1})\\
-\int  d\l Q(\l,t_{4})Q(\l,t_{1})\left[1\pm Q(\l,t_{3})Q(\l,t_{2})\right]\rho(\l|Q,t_{1})
\end{multline}
This is just an algebraic step. Now take the modulus of both sides and use the triangle inequality to get,
\begin{multline}
\left|\ex{\hat Q(t_{2})\hat Q(t_{1})}-\ex{\hat Q(t_{4})\hat Q(t_{1})}\right|\\
=\int  d\l \left[1\pm Q(\l,t_{4})Q(\l,t_{3})\right]\rho(\l|Q,t_{1})\\
+\int  d\l \left[1\pm Q(\l,t_{3})Q(\l,t_{2})\right]\rho(\l|Q,t_{1})\\
\shoveleft{\left|\ex{\hat Q(t_{2})\hat Q(t_{1})}-\ex{\hat Q(t_{4})\hat Q(t_{1})}\right|}\\
\leq 2 \pm \left[\int  d\l  Q(\l,t_{4})Q(\l,t_{3})\rho(\l|Q,t_{1})\right.\\
\left.+\int  d\l Q(\l,t_{3})Q(\l,t_{2})\rho(\l|Q,t_{1})\right]\label{2.8}
\end{multline}
Again this is simply algebraic. However we would now like to identify the averages on the right hand side of Eq.(\ref{2.8}) with some observable quantities. This proves to be impossible because they don't have the correct structure to be two-time correlation functions. What is wrong is that the probability distributions over the hidden variables are functions of a measurement carried out at $t_{1}$, which they should not be, and are not functions of measurements at $t_{2}$ or $t_{3}$, which they should be. In order to remedy this we need to assume that the measurements at $t_{1},t_{2}$ and $t_{3}$ are non-invasive, in the sense of Eq.(\ref{2.3}). This then lets us write,
\begin{multline}
\left|\ex{\hat Q(t_{2})\hat Q(t_{1})}-\ex{\hat Q(t_{4})\hat Q(t_{1})}\right|\\
\leq2\pm \left[\ex{\hat Q(t_{3})\hat Q(t_{2})}+\ex{\hat Q(t_{4})\hat Q(t_{3})}\right],
\end{multline}
which is a Leggett-Garg inequality. 

The derivation above shows that, mathematically at least, the assumption of non-invasive measurability is necessary in order to derive the Leggett-Garg inequalities. 

\section{A Realistic Hidden Variables Theory Which Violates The Leggett-Garg Inequality}

In this section we wish to demonstrate that is is possible to construct a hidden variables theory which realist but does not satisfy the condition of non-invasive measurability and thus can violate the Leggett-Garg inequalities. Let us begin however by pointing out that such a theory already exists, in the form of de Broglie-Bohm theory \cite{dBB}. Thus we could simply quote this theory and end our argument. However there are at least two good reasons for not doing this. The first is that de Broglie-Bohm theory, whilst empirically adequate, is rather cumbersome when it comes to handling finite dimensional quantum systems such as the two level oscillator usually used to demonstrate violation of the Leggett-Garg inequalities. The second reason is that in de Broglie-Bohm theory the invasiveness of the measurements comes about because the full description of the system also includes the state of the measuring device. However this masks the fact that it is possible to model the effects of the measurements as changes to the internal state of the system, rather than as being associated with a specific measuring device. 

\subsection{Standard Quantum Description}

Let us first recall the standard quantum description of this process. We have a quantum state,
\beq
\ket{\psi}=   \vvec{\psi_{1}}{\psi_{2}}
\eeq 
evolving under a Hamiltionan,
\beq
\hat H=\frac{\w}{2}\begin{bmatrix}0&1\\1&0\end{bmatrix}
\eeq
and thus 
\beq
U(t)=e^{-i\hat Ht}=\cos(\w t/2)\begin{bmatrix}1&0\\0&1\end{bmatrix}-i\sin(\w t/2)\begin{bmatrix}0&1\\1&0\end{bmatrix}
\eeq
Measurements on the system are modeled by projection operators,
\beq
\hat P_{1}=\begin{bmatrix}1&0\\0&0\end{bmatrix} \quad \mbox{and } \hat P_{2}=\begin{bmatrix}0&0\\0&1\end{bmatrix}
\eeq
Thus, for an initial state $\ket{\psi_{0}}=\vvec{1}{0}$ the probabilites that a measurement at a time t finds the system in state 1 or 2 are given by,
\bea
p(1,t)&=&\frac{1}{2}(1+\cos(\w t))\\
p(2,t)&=&\frac{1}{2}(1-\cos(\w t))
\eea
and thus 
\beq
\ex{\hat Q(t)}=\cos(\w t).
\eeq
Furthermore, the probabilities for the outcomes of measurements at times $t_{1}$ and $t_{2}$ are;
\bea
p(1,t_{1},1,t_{2})&=&||\hat P_{1}e^{-i\hat H(t_{2}-t_{1})}\hat P_{1}e^{-i\hat Ht_{1}}\ket{\psi_{0}}||^{2}\nonumber\\
&=&\frac{1}{4}(1+\cos(\w(t_{2}-t_{1})))(1+\cos(\w t_{1})),\nonumber\\
p(1,t_{1},2,t_{2})&=&\frac{1}{4}(1-\cos(\w(t_{2}-t_{1})))(1+\cos(\w t_{1})),\nonumber\\
p(2,t_{1},1,t_{2})&=&\frac{1}{4}(1-\cos(\w(t_{2}-t_{1})))(1-\cos(\w t_{1})),\nonumber\\
p(2,t_{1},2,t_{2})&=&\frac{1}{4}(1+\cos(\w(t_{2}-t_{1})))(1-\cos(\w t_{1})).\nonumber\\
\eea
Notice that the probabilities for two time measurements have the form of products of single time probabilities. This is a consequence of the collapse postulate in standard quantum theory, which effectively `resets' the state after either of the measurements $\hat P_{1/2}$.

We therefore find,
\bea
L&=&\left<\hat Q(t_1)\hat Q(t_2)\right>+\left<\hat Q(t_2)\hat Q(t_3)\right>+\left<\hat Q(t_3)\hat Q(t_4)\right>\\\nonumber
&&-\left<\hat Q(t_1)\hat Q(t_4)\right>\nonumber\\
&=&\cos[\w(t_2-t_1)]+\cos[\w(t_3-t_2)]\nonumber\\
&&+\cos[\w(t_4-t_3)]-\cos[\w(t_4-t_1)]
\eea
Taking $t_1=t$, $t_2=2t$ etc we find,
\beq
L[t]=3\cos[\w t]-\cos[3\w t]\label{3.11}
\eeq
which for $t=\pi/4\w$ is equal to $2\sqrt{2}$, and thus we see this Leggett-Garg inequality is violated for this system.

\subsection{Realist Model}

The realist model we will present is the simplest we can think of that gives agreement with quantum theory. This model was inspired by de Broglie-Bohm theory, and we can think of this as a combined system plus measuring device. Alternatively we can think of this as representing a classical system with a single classical bit of memory. In fact as we shall present it this model works only for a quantum system measured once, subsequent measurements tend to disturb the system in such a way as to give rise to behavior not equivalent to quantum theory. If we want we can rectify this by extending the model to allow for more bits of memory, but this simpler version is sufficient to demonstrate the basic ideas.

The model has as its state space four different regions labeled by $a,b,c,d$. The hidden variables are the possible trajectories of the system as it moves between the different regions. Free evolution causes oscillations between regions $a$ and $b$ and regions $c$ and $d$. Our measurement will cause a transition between regions $a$ and $c$. 

\begin{figure}[ht!]
     \begin{center}
        \subfigure[ Free evolution causes transitions between states $\psi_{a/b}$ and $\psi_{c/d}$.]{%
            \label{fig:first}
            \includegraphics[width=0.4\textwidth]{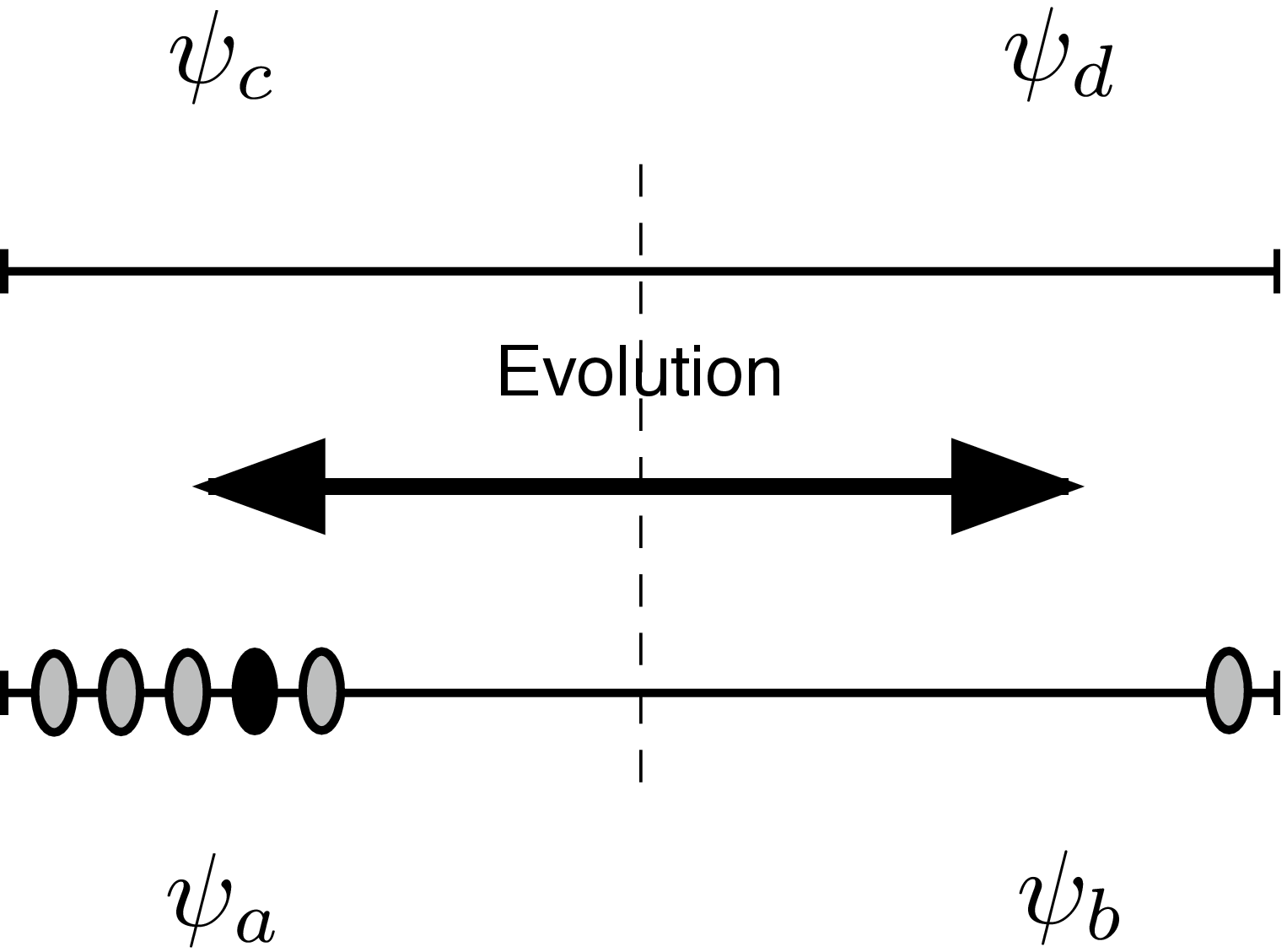}
        }\\%
        \subfigure[As well as reading out the proportion of the state in $\psi_{a}+\psi_{c}$ and $\psi_{b}+\psi_{d}$ measurement of the system also swaps trajectories in $\psi_{a/c}$.  ]{%
           \label{fig:second}
           \includegraphics[width=0.4\textwidth]{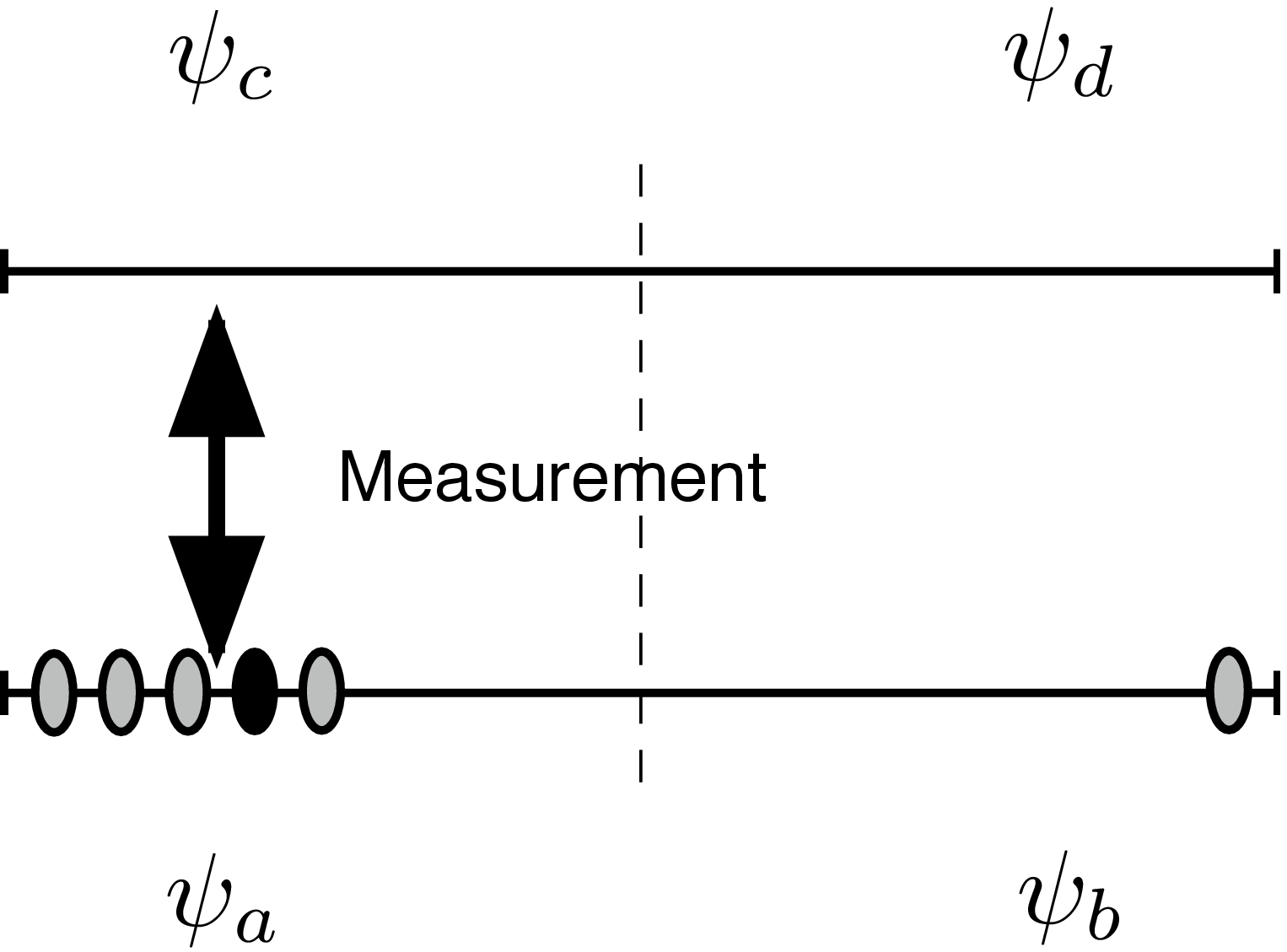}
        }
    \end{center}
    \caption{%
        Pictorial representation of our hidden variable model. The dynamics comes in two parts, firstly a free evolution and secondly a disturbance due to measurement.
     }%
   \label{fig:subfigures}
\end{figure}

We denote by $\psi_{a}$ the probability of the system being in state $a$ etc and $\psi_{a}+\psi_{b}+...=1$.   What in the quantum system was state 1 maps here onto $\psi_{a}$ and $\psi_{c}$ and likewise for state 2. 

Between measurements the rate at which trajectories cross from $\psi_{a}$ to $\psi_{b}$ is determined by,
\beq
\frac{d^{2}}{d t^{2}}(\psi_a-\psi_b)=-\w^2 (\psi_a-\psi_b)
\eeq 
\beq
\frac{d}{dt}(\psi_a+\psi_b)=0
\eeq
and similarly for $c$ and $d$, together with the condition that the trajectories representing the hidden variables do not cross or coalesce. (This is the analogue of the fact that the guidance equation in de Brogle-Bohm theory is first order in $t$.) One can think of the trajectories as like those of beads on an abacus, which may slide back and forwards between sides but which retain their original ordering. 

Measurements have two different parts. Firstly a measurement of the variable $Q(\l,t)$ returns the value $(\psi_{a}+\psi_{c})-(\psi_{b}+\psi_{d})$, and secondly this measurement causes the trajectories in $\psi_{a}$ to swap with those in $\psi_{c}$. Notice that, as advertised, this model will run into problems after a second measurement. If we desire we can extend this model by incorporating extra bits of memory to take account of subsequent measurements.

Choosing the same initial state as for the quantum system described above means setting $\psi_{a}+\psi_{c}=1$. In addition we will choose $\psi_{a}=1$ but otherwise the choice of which trajectory is realised is assumed to be random with uniform probability. It is easy to see that the probability of finding the system in the state $\psi_{a}$ at a subsequent time $t$ is given by,
\beq
\psi_{a}(t)=\frac{1}{2}(1+\cos(\w t))
\eeq
Noting that the probability of finding the quantum system in state 1 is given by the probability of finding it in states $\psi_{a}$ or $\psi_{c}$ in the hidden variable model we see that,
\beq
p(1,t)=p(a+c,t)=\psi_{a}(t)+\psi_{c}(t)=\frac{1}{2}(1+\cos(\w t))
\eeq
 and thus this model matches the single time predictions for quantum theory. Note that the effect of the measurement on the hidden variables here can be neglected.

Furthermore, the no-crossing property of the trajectories means that the two time probabilities, in the absence of measurement, can be easily shown to be,
\bea
p(a,t_{1},a,t_{2})&=&\min[p(a,t_{1}),p(a,t_{2})],\nonumber\\
p(1,t_{1},2,t_{2})&=&\max[p(a,t_{1})-p(a,t_{2}),0],\nonumber\\
p(2,t_{1},1,t_{2})&=&\max[p(b,t_{1})-p(b,t_{2}),0],\nonumber\\
p(2,t_{1},2,t_{2})&=&\min[p(b,t_{1}),p(b,t_{2})].\label{3.16}
\eea
It can be shown algebraically that these set of two time probabilities give rise to correlation functions which do not violate the Leggett-Garg inequalities, but it is simpler to demonstrate this by taking $t_{1}=t$, $t_{2}=2t$ etc as for the quantum case and plotting the results for various values of $t$. 

\begin{figure}[htbp] 
   \centering
   \includegraphics[width=3.5in]{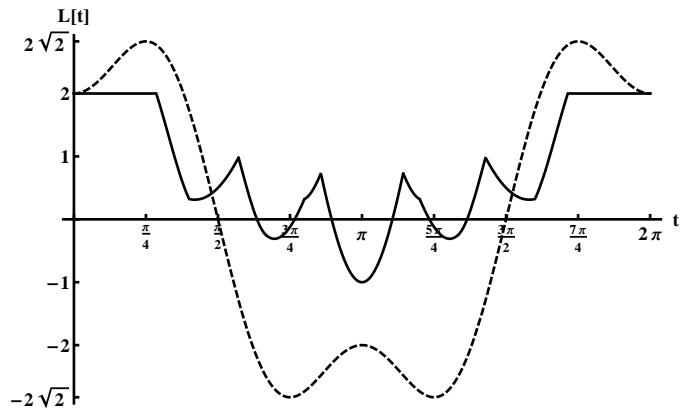} 
   \caption{The value of the quantity $L$ in the Leggett-Garg inequality as given by quantum theory Eq.(\ref{3.11}), dashed line, and computed from our hidden variable theory via Eq.(\ref{3.16}), solid line. The quantum prediction clearly violates the inequality $2\leq L\leq2$.}
   \label{fig:example}
\end{figure}

Now we turn to the question of what happens when we include the effects of measurement. When we measure $Q$ at $t_{1}$ we send the trajectories in $a$ to $c$. Therefore, the values of $Q$ at time $t_{2}$ given that the state was in either $\psi_{a}$ or $\psi_{b}$ at time $t_{1}$ are given by.
\bea
Q(t_{2}|\psi_{a},t_{1})&=& \cos(\w(t_{2}-t_{1})),\nonumber\\
Q(t_{2}|\psi_{b},t_{1})&=& -\cos(\w(t_{2}-t_{1})),\nonumber
\eea
Thus our hidden variable model predicts that the measured value of the correlation functions should be,
\bea
\ex{\hat Q(t_{2})\hat Q(t_{1})}&=&\frac{1}{2}(1-\cos(\w t_{1}))\cos(\w(t_{2}-t_{1}))\nonumber\\
&&+\frac{1}{2}(1+\cos(\w t_{1}))\cos(\w(t_{2}-t_{1}))\nonumber\\
&=&\cos(\w(t_{2}-t_{1}))
\eea
and we reproduce the predictions of the quantum model.

We see therefore that this realistic model does not violate the Leggett-Garg inequalities in the absence of measurement but it does so when we include the effects of measurement. However in the absence of measurement nothing is measured! Thus the fact that this system obeys these inequalities is hidden from us because of the disturbing nature of our measurement of $\hat Q$.

\section{Undetectable vs Non-Invasive Measurements}

Let us turn finally to the question of whether one can demonstrate that some particular experimental set up is performing a non-invasive measurement. Recall that the mathematical expression of this is the condition,
\beq
\rho(\l|A,t_{0})=\rho(\l)
\eeq
where for simplicity we will assume we are dealing only with a single measurement. Firstly let us note that this expression only makes sense in the context of an appropriate hidden variables theory specifying the set $\{\l\}$ complete with dynamics and a method for assigning a probability distribution on the $\l$ given an initial quantum state. It is not a condition that can be directly formulated within quantum theory. 

Suppose instead we are given two sets of a large number of copies of two states of a quantum system and told that one set has previously been subject to a measurement of some observable $\hat A$ at time $t_{0}$. Can we determine, at least with high probability, which set of states has been measured? If we cannot then we might be justified in calling such a measurement `undetectable.' The key question is, is such a measurement necessarily `non-invasive'? Let us denote the unmeasured and measured states in the quantum and hidden variable theories as $\psi\equiv\rho(\l)$ and $\psi'\equiv\rho(\l|A,t_{0})$ respectively. Suppose it is the case that for all subsequent measurements $\hat B(t)$ that we can perform,
\begin{multline}
\ex{\hat B(t)}_{\psi}=\int d\l B(\l,t)\rho(\l)\\
=\ex{\hat B(t)}_{\psi'}=\int d\l B(\l,t)\rho(\l|A,t_{0}).
\end{multline}
In this case the measurement at $t_{0}$ is an extreme example of an undetectable measurement since there is no subsequent measurement that can distinguish the two states. Technical issues notwithstanding, we can infer from this that $\psi=\psi'$. 

But is this measurement therefore non-invasive? {\em ie} does the fact that $\psi=\psi'$ imply that $\rho(\l)=\rho(\l|A,t_{0})$? The answer depends on the properties of the hidden variable theory we are considering. Hidden variable theories may be grouped into two classes, usually called {\em psi-ontic} and {\em psi-epistemic} \cite{SpHa}. The distinction is essentially over whether the quantum state of the system over or under specifies the distribution of hidden variables. Theories where the mapping between quantum states and distributions of hidden variables is many-to-one are called {\em psi-epistemic} and theories where the mapping is one-to-many are called {\em psi-ontic}. Recent results such as the PBR theorem \cite{PBR} tend to rule out hidden variable theories of the psi-epistemic type, but for the moment we shall consider both on an equal footing.

Let us suppose our hidden variable theory is of the psi-epistemic type. Then there are three possible types of measurement;  measurements which change the quantum state and the ontic state, measurements which change the quantum state but not the ontic state, and measurements which change neither the quantum state nor the ontic state.
However suppose our hidden variable theory is of of the psi-ontic type. Again there are three possible types of measurements; measurements which change the quantum state and the ontic state, measurements which change the ontic state but not the quantum state, and measurements which change neither the quantum state nor the ontic state. 

For a psi-epistemic theory all measurements which do not change the quantum state cannot change the ontic state, and thus for psi-epistemic theories all undetectable measurements are also non-invasive measurements. However for psi-ontic theories this is not the case. Psi-ontic theories allow for the existence of non-invasive measurements, but it is impossible to prove from any number of subsequent measurements carried out on a quantum system that a given prior measurement was non-invasive.

\section{Summary}

In this paper we have given an introduction to the Leggett-Garg inequalities hopefully suitable for students and researchers new to the field. We began by giving a derivation of the inequalities which emphasizes the role of the assumption of non-invasive measurability. We then gave an example of a hidden variable theory which is realist but not non-invasively measurable and showed that this could reproduce the predictions of quantum theory, and in particular could account for violations of the Leggett-Garg inequalities. We finally turned to the question of whether non-invasive measurability could be experimentally established for a given measurement set up. We pointed out that in general it is impossible to prove non-invasive measurability at the level of the hidden variables from any quantum measurements and that in particular `undetectable measurements' need not be non-invasive.

We have shown that in general it is impossible to prove non-invasive measurability from the results of any subsequent measurements performed on a system. Nevertheless it may be possible to make progress if we are willing to restrict the class of possible hidden variable theories in some way. For example limiting the allowed number of bits of memory in a hidden variable theory, or equivalently the dimension of its state space, may be enough to rule out realist models reproducing certain features of quantum theory. See \cite{LGextraextra} for related ideas. Alternatively one may imagine restricting attention to hidden variable theories where the disturbance due to measurement takes a particular form. These ideas are unlikely to be of much value in the context of pure quantum theory, but they might prove useful in situations where quantum behavior is only emergent from a deeper dynamics, and where we can thus constrain in some way the underlying hidden variable thory. An important example of this is the quantum modeling of cognitive processes in psychology \cite{BuBr}. These ideas will be pursued elsewhere \cite{PoYe}.

\acknowledgments

This research was supported by a grant from the John Templeton Foundation. The author would also like to thank EM Pothos  and MJS Lee for useful conversations on this topic.

\end{document}